\documentclass[12pt,preprint]{aastex}

\shorttitle{Homologous Gravitational Collapse}
\shortauthors{Tsui and Navia}

\begin{document}

\title{Homologous Gravitational Collapse in Lagrangian Coordinate:
\\
Planetary System in Protostar and Cavity in Pre-Supernova}
\author{K.H. Tsui and C.E. Navia}
\affil{Instituto de F\'{i}sica - Universidade Federal Fluminense
\\Campus da Praia Vermelha, Av. General Milton Tavares de Souza s/n
\\Gragoat\'{a}, 24.210-346, Niter\'{o}i, Rio de Janeiro, Brasil.}
\email{tsui$@$if.uff.br}
\pagestyle{myheadings}
\baselineskip 18pt
	 
\begin{abstract}

The classical problem of spherical homologous gravitational collapse
 with a polytropic equation of state for pressure
 is examined in Lagrangian fluid coordinate,
 where the position of each initial fluid element
 $\eta=r(0)$ is followed in time by the evolution function $y(t)$.
 In this Lagrangian description,
 the fluid velocity $v=dr/dt=\eta dy/dt$ is not a fluid variable,
 contrary to the commonly used Eulerian fluid description.
 As a result, the parameter space is one dimensional in $\eta$,
 in contrast to the $(x,v)$ two-parameter space
 of Eulerian formulation.
 In terms of Lagrangian coordinate,
 the evolution function $y(t)$,
 which is not limited to a linear time scaling,
 agrees with the well established parametric form
 of Mestel \citep{mestel1965} for cold cloud collapse.
 The spatial structure is described by an equation
 which corresponds to the one derived by
 Goldreich and Weber \citep{goldreich1980}.
 The continuous self-similar density distribution
 presents a peaked central core
 followed by oscillations with decreasing amplitude,
 somewhat reminiscent to the expansion-wave
 inside-out collapse of Shu \citep{shu1977}.
 This continuous solution could account for
 the planetary system of a protostar.
 There is also a disconnected density distribution,
 which could be relevant to cavity formation
 between the highly peaked central core
 and the external infalling envelope
 of a magnetar-in-a-cavity pre-supernova configuration.

\end{abstract}
\keywords{gravitational collapse}

\maketitle

\newpage
\section{Gravitational Collapse}

The fundamental issue of gravitational collapse
 of a large but finite massive gas cloud
 concerns the star and nebula formation.
 Bonner \citep{bonner1956} had demonstrated
 that the equilibrium of a finite isothermal
 gas cloud under its self-gravity
 could be unstable as the mass increases.
 Mestel \citep{mestel1965}
 solved the collapse equation of a cold cloud
 with $v=dr/dt$ as the trajectory velocity,
 which is the Lagrangian fluid representation,
 to obtain the celebrated self-similar parametric solution.
 Lin \citep{lin1965} showed
 that an oblate cold spheroid would evolve to a disk
 and a prolate one would evolve to a spindle.
 Bodenheimer and Sweigart \citep{bodenheimer1968}
 studied numerically the evolution sequence
 of a collapsing gas cloud with finite preesure
 again with $v=dr/dt$
 under different initial density distributions
 and different surface boundary conditions.
 Penston \citep{penston1969} analyzed analytically
 the Lagrangian cold collapse with a smooth maximum for density,
 and also put forward a self-similar analysis
 of an isothermal collapsing sphere in Eulerian description,
 where the fluid velocity $v$ as one of the variables.
 Larson \citep{larson1969} examined numerically
 through a set of conservation equations
 of mass, momentum, and energy in Eulerian representation
 the formation of a protostar,
 and presented in appendix C the isothermal similarity solutions.
 Shu \citep{shu1977} studied anew
 the homologous collapse of an isothermal sphere
 with the Eulerian conservation equations.
 He interpreted the singular solutions,
 where the coefficients of the nonlinear
 differential equations vanish,
 and constructed the expansion-wave,
 inside-out collapse scenario.
 Hunter \citep{hunter1977} added a new class
 of isothermal self-similar solutions
 on previously known ones.
 Goldreich and Weber \citep{goldreich1980}
 examined the homologous collapse of a stellar core
 with a polytropic equation of state,
 and arrived at a differential equation
 that describes the radial structure.
 Perturbation analysis was used to study
 the nonspherical modes of the stellar core.
 Whitworth and Summers \citep{whitworth1985}
 brought to the attention the importance of
 the stability of the initial isothermal gas cloud
 and the external driving pressure.
 These two factors transform
 each known solution into a continum.

Recently, there have been renewed efforts
 on gravitational collapse.
 Fatuzzo \citep{fatuzzo2004},
 Lou and Wang \citep{louwang2006},
 Lou and Gao \citep{lou2006}
 have considered self-similar solutions
 in Eulerian form with polytropic equation of state,
 and presented innovated solutions
 for astrophysical phenomena.
 Furthermore, Lou and Shen \citep{lou2004}
 put forward the envelope-expansion (EE)
 and core-collapse (CC) solution for an isothermal cloud.
 This EECC solution is to model outflowing stellar winds
 while the red giant collapses.
 Bian and Lou \citep{bian2005} examined the shock flows
 of an isothermal sphere
 to better understand the shock structure.

We remark that, except cold gas clouds,
 clouds with isothermal or polytropic equation of state
 have been treated under Eulerian fluid description
 where velocity $v=v(r,t)$ is one of the variables.
 The time dependence of each variable
 is constructed as some specific power of time $t$.
 The spatial dependence is given by
 a set of nonlinear differential equations \citep{lou2004}.
 The singular solutions of this system
 define the singular sonic surfaces,
 which separate subsonic and supersonic regions
 in the $(x,v)$ two-parameter space.
 Along the sonic surfaces,
 the flow is subsonic/supersonic
 relative to the similarity $x$ profile.
 Collapse solutions cross from
 subsonic to supersonic regions
 along the sonic lines.
 Sometimes shock waves are needed
 to bridge the crossing
 to meet the boundary conditions.
 Aimed to describe the collapse,
 representing the time dependence
 with different powers of a linear time scaling
 for different variables
 would certainly fail to describe
 the time history of collapse.
 We could only expect such self-similar solutions
 with a linear time scaling
 be valid during a specific time span
 and over a specific radial interval.
 Here, we use the alternative Lagrangian description,
 where the fluid velocity is not a variable,
 to construct the self-similar collapse
 having a polytropic equation of state.
 The evolution function is solved consistently
 together with the spatial structures.
 
\newpage
\section{Lagrangian Similarity}

Gravitational collapse of a spherical cloud is described by
\begin{eqnarray}
{\partial\rho\over\partial t}+\nabla\cdot\left(\rho\vec v\right)\,&
 =&\,0\,\,\,,\label{eqno1}
\\
\rho\left\{{\partial\vec v\over\partial t}
 +(\vec v\cdot\nabla)\vec v\right\}\,&
 =&\,-\nabla p-\rho{GM_{*}\over r^2} \hat r
 -\rho{GM(r,t)\over r^2} \hat r
 \,\,\,,\label{eqno2}
\\
{\partial\over\partial t}\left({p\over\rho^{\gamma}}\right)
 +(\vec v\cdot\nabla)\left({p\over\rho^{\gamma}}\right)\,&
 =&\,0\,\,\,.\label{eqno3}
\end{eqnarray}
Here, $\rho$ is the mass density, $\vec v$ is the gas cloud velocity,
 $p$ is the gas pressure, $\gamma$ is the polytropic index,
 $M_{*}$ is a point mass at the center, if any,
 and $M(r,t)$ is the gas mass within a sphere of radius $r$ at time $t$
 where
\begin{eqnarray}
M(r,t)\,=\,\int_{0}^{r} 4\pi r'^{2}\rho(r',t)dr'\,\,\,.\label{eqno4}
\end{eqnarray}
We note that Eq.~\ref{eqno1} can be written alternatively as
\begin{eqnarray}
{\partial M\over\partial t}+v{\partial M\over\partial r}\,
 =\,C(t)\,=\,0\,\,\,,\label{eqno5}
\end{eqnarray}
where $C(t)$ is a constant independent of $r$
 and varies only in time.
 By taking $C(t)=0$, we would have an inhomogeneous cloud
 that is stationary in time.
 For a spherical collapse with $\vec v=v\hat r$,
 we use Lagrangian fluid representation
 with $r(t)=r(0)y(t)=\eta y(t)$,
 where the Lagrangian label $\eta$
 is the initial position of the fluid element,
 and $y(t)$ is the dimensionless evolution function with $y(0)=1$
 that describes the time history of the fluid element.
 The velocity can then be written as
\begin{equation}
v\,=\,\eta{dy\over dt}\,\,\,.\label{eqno6}
\end{equation}
Contrary to the Eulerian fluid description,
 the fluid velocity $v$ here is not a variable
 which substantially simplifies the analysis.

We now transform the Eulerian independent variables $(r(t),t)$
 to Lagrangian independent variables $(\eta,y(t))$.
 Furthermore, we note that the total time derivative
 in Eulerian representation
 $(\partial/\partial t+v\partial/\partial r)$
 corresponds to the partial time derivative
 in Lagrangian representation $\partial/\partial t$.
 We now write the variables in saparable form of $y$ and $\eta$,
 and try to determine the evolution function $y(t)$
 and the spatial configuration in $\eta$ under similarity.
 Following Tsui and Serbeto \citep{tsui2007},
 we have from Eq.~\ref{eqno5} and Eq.~\ref{eqno3}
\begin{eqnarray}
M(\eta,y)\,=\,M_{0}{1\over y^{0}}\bar M(\eta)\,\,\,,\label{eqno7}
\\
F\,=\,{p\over\rho^{\gamma}}\,
 =\,{p_{0}\over\rho_{0}^{\gamma}}
 {1\over y^{0}}\bar F(\eta)\,\,\,,\label{eqno8}
\end{eqnarray}
where $M_{0}$ carries the physical dimension of $M(\eta,y)$
 such that $\bar M(\eta)$ is dimensionless,
 and likewise are $p_{0}/\rho_{0}^{\gamma}$ and $\bar F(\eta)$.
 We note that $F$ is only a function of entropy
 which remains constant in time and in space.
 For this reason, we have $\bar F(\eta)=1$.
 Besides, from Eq.~\ref{eqno4}
 and with $\rho(\eta,y)=\rho_{0}Y(y)\bar\rho(\eta)$,
 we have
\begin{eqnarray}
M(r,t)\,=\,y^{3}\int_{0}^{\eta} 4\pi \eta'^{2}\rho_{0}Y(y)\bar\rho(\eta')d\eta'\,\,\,.\label{eqno9}
\end{eqnarray}
Comparing with Eq.~\ref{eqno7} gives
\begin{eqnarray}
\rho(\eta,y)\,=\,\rho_{0}{1\over y^{3}}\bar\rho(\eta)\,\,\,,\label{eqno10}
\\
\bar M(z)\,=\,\int_{0}^{z}3z'^{2}\bar\rho(z')dz'\,\,\,,\label{eqno11}
\\
M_{0}\,=\,{4\pi\over 3}{\rho_{0}\over a^{3}}\,\,\,,\label{eqno12}
\end{eqnarray}
where we have introduced the scaling $a$ to write $\eta$
 in normalized form $z=a\eta$.
 Making use of Eq.~\ref{eqno8}, we now have
\begin{eqnarray}
p(\eta,y)\,
 =\,p_{0}{1\over y^{3\gamma}}\bar\rho^{\gamma}(\eta)\,\,\,.\label{eqno13}
\end{eqnarray}
As for Eq.~\ref{eqno2}, in terms of Lagrangian coordinate
 and with $\gamma=4/3$, it reads
\begin{eqnarray}
y^{2}{d^{2}y\over dt^{2}}\,
 =\,-{p_{0}a^{2}\over\rho_{0}}{1\over\bar\rho}{1\over z}
 {\partial\bar p\over\partial z}
 -{GM_{*}a^{3}\over z^{3}}
 -{GM_{0}a^{3}\over z^{3}}\bar M\,
 =\,C\,\,\,.\label{eqno14}
\end{eqnarray}
This equation has the temporal and spatial parts separated
 with $C$ as the separation constant.
 
\newpage
\section{Evolution Function}

Choosing $C=-NGM_{0}a^{3}$ in Eq.~\ref{eqno14}
 and defining $\tau=(GM_{0}a^{3})^{1/2}t=\omega t$,
 the temporal part of Eq.~\ref{eqno14} is
\begin{eqnarray}
{d^{2}y\over d\tau^{2}}\,
 =\,-{N\over y^{2}}\,\,\,.\label{eqno15}
\end{eqnarray}
Taking $N=0$ would give $dy/d\tau=-n$
 and $y(\tau)=y(0)-n\tau$,
 which reproduces the linear time scaling
 of the Eulerian formulation.
 For $n=0$, we have the equilibrium state
 of a gravitating shherical cloud.
 For $N>0$, $y(\tau)$ has a downward curvature.
 This is compatible to an inflowing fluid to the center
 with $y(\tau)$ decreasing from its initial value of $y(0)=1$.
 The first integral of this equation is
\begin{eqnarray}
\left({dy\over d\tau}\right)^{2}+\left(-{2N\over y}\right)\,
 =\,-H\,\,\,.\label{eqno16}
\end{eqnarray}
The terms on the left side could be interpreted
 as the kinetic and potential energies
 and the constant $-H$ on the right side
 could be regarded as the total energy.
 The inflow rate is given by
\begin{eqnarray}
{dy\over d\tau}\,
 =\,\pm\left[\left({2N\over y}-H\right)\right]^{1/2}\,
 =\,-\left[\left({2N\over y}-H\right)\right]^{1/2}\,\,\,.\label{eqno17}
\end{eqnarray}
With $H=2N$, the initial velocity $dy/dt=0$ starts from zero,
 and Fig.1 shows the evolution function $y(\tau)$
 and its time derivative $dy/d\tau$ respectively.

We remark that this evolution function
 actually corresponds to the parametric time function
 of Mestel \citep{mestel1965}.
 Taking $y=\cos^{2}\theta$, Eq.~\ref{eqno16} with $2N=H$ gives
\begin{eqnarray}
\left({dy\over d\tau}\right)^{2}\,
 =\,H\tan^{2}\theta\,\,\,.\nonumber
\end{eqnarray}
With $dy/d\tau=2\cos\theta\sin\theta d\theta/d\tau$,
 we have
\begin{eqnarray}
2\cos^{2}\theta d\theta\,
 =\,(1+\cos2\theta)d\theta\,
 =\,H^{1/2}d\tau\,\,\,.\nonumber
\end{eqnarray}
Integrating once then gives
\begin{eqnarray}
\theta+{1\over 2}\sin2\theta\,
 =\,H^{1/2}\tau\,\,\,,\label{eqno18}
\end{eqnarray}
which is the well established parametric solution of Mestel.

\newpage
\section{Spatial Structure}

With $C=-NGM_{0}a^{3}$, the spatial part of Eq.~\ref{eqno14} is
\begin{eqnarray}
{p_{0}a^{2}\over\rho_{0}}
 {1\over GM_{0}a^{3}}
 {1\over\bar\rho}{1\over z}
 {\partial\bar p\over\partial z}
 +{\bar M\over z^{3}}\,
 =\,N\,\,\,.\label{eqno19}
\end{eqnarray}
With $\bar p=\bar\rho^{\gamma}$,
 $C^{2}_{s0}=\gamma p_{0}/\rho_{0}$,
 and $v^{2}_{ff}=2GM_{0}a$ where $ff$ denotes free fall,
 we get
\begin{eqnarray}
{C^{2}_{s0}\over v^{2}_{ff}}{2\over (\gamma-1)}
 z^{2}{\partial\over\partial z}\bar\rho^{\gamma-1}\,
 =\,-\int^{z}_{0}\bar\rho(z')3z'^{2}dz'
 +Nz^{3}\,\,\,.\label{eqno20}
\end{eqnarray}
Denoting $\alpha=C^{2}_{s0}/v^{2}_{ff}$,
 we get
\begin{eqnarray}
2\alpha z^{2}{\partial\over\partial z}\bar\rho^{\gamma-1}\,
 =\,-3(\gamma-1)\int^{z}_{0}(\bar\rho(z')-N)z'^{2}dz'
 \,\,\,.\label{eqno21}
\end{eqnarray}
With $\gamma=4/3$, and writing $\bar\rho^{1/3}=q$
 and $\bar\rho=q^{3}$, Eq.~\ref{eqno21} becomes
\begin{eqnarray}
2\alpha z^{2}{\partial q\over\partial z}\,
 =\,-\int^{z}_{0}(q^{3}(z')-N)z'^{2}dz'
 \,\,\,.\label{eqno22}
\end{eqnarray}
Differentiating this equation once
 leads to the following equation,
 which is similar to Eq.6
 of Goldreich and Weber \citep{goldreich1980},
\begin{eqnarray}
2\alpha{1\over z^{2}}
 {\partial\over\partial z}
 \left(z^{2}{\partial q\over\partial z}\right)
 +q^{3}\,
 =\,N\,\,\,.\label{eqno23}
\end{eqnarray}
In Goldreich and Weber,
 $N$ is their free parameter $\lambda$
 which approaches zero,
 and $q(0)$ is chosen as unity.
 In our case, $\alpha$ and $q(0)$
 are the free parameters.
 We remark that Bodenheimer and Sweigart \citep{bodenheimer1968}
 had pointed out the importance of the parameter $\alpha$.
 The most important difference
 is our constant $N=1$,
 which comes from the separation constant $C$.
 Should we take $C=0$ and thus $N=0$,
 we would have $dy/d\tau=-n$ and $y=y(0)-n\tau$
 from Eq.~\ref{eqno15}.
 This would recover the linear time scaling
 of the Eulerian similarity
 and Eq.~\ref{eqno23} would equal to
 Eq.6 of Goldreich and Weber.
 The fact that $N=1$ not only
 provides an evolution function of Mestel,
 but also changes qualitatively
 the nature of spatial solutions.

With $\alpha=0$ for a cold fluid,
 the equilibrium requires $dy/d\tau=0$ with $y(\tau)=y(0)=1$.
 This means $N=0$ and $n=0$,
 and it is obvious from Eq.~\ref{eqno22}
 that such a cold fluid equilibrium does not exist,
 other than $q(z)=0$.
 As for the collapse solution of $N>0$,
 we have the uniform density sphere solution of 
\begin{eqnarray}
\bar\rho(z)\,=\,q^{3}(z)\,=N\,\,\,.\nonumber
\end{eqnarray}
This states that a uniform density cold sphere
 under homologous infall stays as a uniform sphere.
 By taking $N=1$, $\rho_{0}$ amounts to the density of the sphere.
 Making use of Eq.~\ref{eqno11}, $\bar M(z_{max})=z^{3}_{max}$,
 and therefore $M_{0}z^{3}_{max}$ is the mass
 of the entire uniform sphere of $z_{max}$.

With $\alpha\neq 0$, equilibrium configuration with $N=0$
 gives a monotonically decreasing profile
 that ends at a finite $z_{max}<\infty$ with $q(z_{max})=0$.
 With $q(0)=2$ and $\alpha=100$,
 the profiles of $\bar\rho=q^{3}$ and $q$
 are shown in Fig.2 with $z_{max}=40$.
 As for the collapse solution,
 we note that when $q>N=1$,
 the integral of Eq.~\ref{eqno22} is positive,
 and $\partial q/\partial z$ is negative with $q$ decreasing.
 As the upper limit $z$ of the integral increases,
 $q$ would decrease to less than unity,
 and the integral would begin to decrease
 making $\partial q/\partial z$ to turn around.
 We would then get oscillating solutions about $q=1$,
 as shown in Fig.3.
 This oscillating structure
 with a progressively smaller amplitude
 shows that the central core collapses faster
 than the external oscillating envelope.
 This spatial structure bears great ressemblance
 to the expansion-wave inside-out collapse
 scenario of Shu \citep{shu1977}.
 For a finite system with $z_{max}<\infty$,
 the solution is a function of $q(0)$ for a given $\alpha$.
 As $q(0)$ increases with the same $\alpha$,
 the first minimum becomes more extensive,
 and followed by damped ripples about $q=N=1$.
 The peak about $q(0)$ corresponds to the central core,
 and the oscillating part amounts to the external envelope.
 
This spherical oscillating envelope structure of $q(z)$
 could be very relevant to protostar planetary system.
 The locations of the gaseous planets
 would be given by the oscillating peaks of $\bar\rho=q^{3}$.
 In the presence of rotation, as collapse proceeds,
 the spherical cloud would flatten to the equatorial plane,
 and the high density shells in the spherical cloud
 would become dense rings in the equatorial disk
 forming gaseous planets.
 Rotation would also stabilize the planets on their orbits
 as the central pre-protostar continues to collapse.
 Since the peaks are decreasing in amplitude,
 the gaseous planets would have a decreasing mass
 as distance increases.
 This happens to be the case of the giant planets
 of our Solar system,
 with the exception of Uranus
 which has its rotational axis on the ecliptic plane
 indicating a possible collision
 with substantial mass loss.

The fact that the boundary condition at $z_{max}$
 for equilibrium with $q(z_{max})=0$
 differs from that for collapse with $q(z_{max})\approx 1$
 should not cause concern.
 Any self-similar solution,
 either Eulerian or Lagrangian,
 relies on writing the temporal and spatial dependences
 of any dynamic variable in separable form.
 This excludes the initial period of evolution of the real system,
 which is not amenable to similarity treatment.
 As a result, the self-similar initial time $\tau=0$
 is not the initial time $t=0$
 of the real collapsing system.
 Thus, the homologous boundary condition at $\tau=0$
 need not be the boundary condition of equilibrium.
 
For $q(0)\gg 1$ and with a smaller $\alpha$,
 the first minimum plunges to negative $q$ as in Fig.4,
 and becomes positive again at $z_{0}=4.5$.
 As a result, the external part becomes disconnected
 physically from the central core,
 and a cavity is formed between the central core
 and the external envelope. 
 The envelope will collapse
 under the gravitational field of the core
 plus the self-gravity.
 Such a configuration is most relevant to pre-supernova
 (magnetar-in-a-cavity model \citep{uzdensky2007}).
 The envelope of this disconnected system is described by
\begin{eqnarray}
2\alpha z^{2}{\partial q\over\partial z}\,
 =\,-(\gamma-1)\left({M_{*}\over M_{0}}+\bar M-1\right)
 \,\,\,,\label{eqno24}
\\
\bar M(z)\,
 =\,\int_{z_{0}}^{z}q^{3}(z')3z'^{2}dz'\,\,\,,\nonumber
\end{eqnarray}
where the integral $\bar M(z)$ starts at a disconnected $z_{0}>0$.
 With $\gamma=4/3$ and neglecting the self-gravity $\bar M(z)$,
 we have the monotonic envelope profile of
\begin{eqnarray}
6\alpha q(z)\,
 =\,6\alpha q(z_{0})
 -\left({M_{*}\over M_{0}}-1\right)
 \left({1\over z_{0}}-{1\over z}\right)
 \,\,\,.\label{eqno25}
\end{eqnarray}

\newpage
\section{Model Parameters and Sonic Surfaces}

This model has two parameters $M_{0}$ and $a$
 to normalize time $t$ and radial coordinate $\eta$,
 that need to be determined.
 To obtain these two important parameters,
 we begin with observations of a collapsed configuration,
 such as Swift \citep{swift2005}.
 Observing the central core dimension,
 we denote it by $r_{1}$
 which separates from the first minimum.
 This position is given by
\begin{eqnarray}
r_{1}(\tau_{end})\,=\,y(\tau_{end})\eta_{1}\,
 =\,y(\tau_{end})r_{1}(0)\,\,\,,\label{eqno26}
\end{eqnarray}
where $\tau_{end}$ is the end time
 of the evolution function
 with $y(\tau_{end})$ vanishingly small, but nonzero.
 This $y(\tau_{end})$ has to be estimated independently.
 From the peak core density
 and the asymptotic envelope density,
 we can get the normalized $q(0)$ from $\bar\rho(0)=q^{3}(0)$.
 With $q(0)$ determined by observations,
 we get the analytic solution of $\bar\rho(z)$
 with a given $\alpha=C^{2}_{s0}/v^{2}_{ff}$.
 We choose $\alpha=\alpha_{1}$
 such that the inflection point $z_{1}=a\eta_{1}=1$
 to get
\begin{eqnarray}
1\,=\,z_{1}\,=\,a\eta_{1}\,=\,ar_{1}(0)\,
 =\,{y(0)\over y(\tau_{end})}r_{1}(\tau_{end})
 \,\,\,.\label{eqno27}
\end{eqnarray}
With $y(0)=1$ and $y(\tau_{end})$ estimated,
 we can construct backwards in time
 the position $r_{1}$ at $\tau=0$ giving
\begin{eqnarray}
a\,=\,{1\over\eta_{1}}\,\,\,.\label{eqno28}
\end{eqnarray}
If the collapsed configuration
 allows the identification of the number
 of oscillations on the envelope,
 we can compare with the analytic $\bar\rho(z)$
 to locate $z_{max}$.
 By estimating the total mass M
 of the collapsed configuration,
 we then have
\begin{eqnarray}
M\,=\,M_{0}z^{3}_{max}\,\,\,,\label{eqno29}
\end{eqnarray}
which gives $M_{0}$ as the mass
 in unit volume of $z^{3}=1$.
 With $M_{0}$ and $a$ determined,
 we go back to the parameter $\alpha=\alpha_{1}$
 to get $C^{2}_{s0}=\alpha_{1}GM_{0}a$.
 We now calculate
\begin{eqnarray}
\bar M(z_{max})\,
 =\,\int_{z_{0}}^{z_{max}}\bar\rho(z')3z'^{2}dz'\,
 =\,z^{3}_{max}\,\,\,.\label{eqno30}
\end{eqnarray}
Since $\bar\rho(z)$ oscillates about unity,
 $\bar M(z)$ needs to be integrated in $z$.
 However, $\bar M(z_{max})$ can be evaluated easily as above,
 since the collapse is merely a redistribution of densities
 of an initially uniform sphere with $\bar\rho(z)=1$,
 thus giving the second equility of the above equation.

In the Eulerian formulation,
 the singular sonic surfaces
 divide the $(x,v)$ two-parameter space
 in subsonic and supersonic regions,
 where the flow is subsonic/supersonic
 relative to the similarity profile \citep{whitworth1985}.
 Complete collapse solutions
 have to cross these regions
 on the sonic lines.
 In our Lagrangian formulation,
 the fluid velocity,
 which is not a variable,
 and the sound speed are
\begin{eqnarray}
v\,=\,\eta{dy\over dt}\,
 =\,(GM_{0}a)^{1/2}z{dy\over d\tau}\,\,\,.\label{eqno31}
\\
C_{s}\,=\,\left(\gamma p\over\rho\right)^{1/2}\,
 =\,\left(\gamma p_{0}\over\rho_{0}\right)^{1/2}q^{1/2}(z)\,
 =\,C_{s0}q^{1/2}(z)\,=\,C_{s0}\bar C_{s}(z)\,\,\,.\label{eqno32}
\end{eqnarray}
We define the time dependent sonic function
\begin{eqnarray}
S(z,\tau)\,=\,{v\over C_{s}}\,
 =\,{1\over (2\alpha)^{1/2}}
 {z\over\bar C_{s}(z)}{dy\over d\tau}\,
 =\,\bar S(z){dy\over d\tau}\,\,\,.\label{eqno33}
\end{eqnarray}
With $q(0)=5$ and $\alpha=5$,
 the spatial part $\bar S(z)$ is shown in Fig.5
 with oscillations in $z$.
 At $\tau=0$ with $dy/d\tau=0$,
 the time dependent sonic function $S(z,\tau)=0$ everywhere.
 As $\tau>0$ and $dy/d\tau$ increases,
 $S(z,\tau)$ rises up and will cross
 the horizontal line of unit value
 at some point $z_{s}$ where $S(z_{s},\tau)=1$.
 This is the time dependent sonic surface $z_{s}(\tau)$.
 As $\tau$ increases, this sonic surface moves
 relative to the similarity profile towards the center.
 Since $dy/d\tau$ is related
 to the fluid velocity $v=\eta dy/d\tau$,
 the movement of $z_{s}(\tau)$
 amounts to the fluid velocity or flow speed
 with respect to the similarity profile
 \citep{whitworth1985}.
 We, therefore, have a picture of sonic surface
 which is compatible to that
 of the Eulerian formulation \citep{louwang2006}.
 Nevertheless, in the Lagrangian formulation,
 $v$ is directly incorporated in the evolution function,
 and the spatial structures are described by Eq.~\ref{eqno22},
 which is analytic everywhere
 in the one dimensional parameter space $\eta$.
 There is no need of sonic lines to cross
 subsonic/supersonic regions.

\newpage
\section{Self-Organization and Conclusions}

Contrary to laboratory self-organized phenomena
 \citep{hasegawa1985},
 astrophysical events given enough time
 with specific patterns of outcome
 are self-organized, by definition.
 Gravitational collapses,
 extragalactic jets \citep{tsui2007},
 planetary nebulae \citep{tsui2008}, and supernovae
 are some of the examples.
 If, and only if,
 self-similar description is able to represent
 the astrophysical configurations in late times,
 it could only mean
 that these self-similar solutions
 are the attractors of dynamic time evolutions
 from a large set of random initial conditions.
 For the specific case of gravitational collapse,
 the traditional Eulerian homologous formulation,
 where the fluid velocity $v$ is one of the variables,
 has the time dependence of each variable
 expressed in a given power of a linear time scaling.
 The spatial dependence is described by
 a set of nonlinear differential equations.
 The sonic surfaces and sonic lines
 are crucial in constructing complete solutions
 in the $(x,v)$ two-parameter space.

We have presented a Lagrangian homologous formulation,
 where the fluid velocity $v$ is not a variable,
 but is derived from the evolution function $y(t)$.
 The fact that $v$ is not a variable is extremely important,
 because the parameter space is now reduced
 to one dimensional in $\eta$.
 In particular, no sonic lines are needed
 to cross subsonic/supersonic regions.
 The crossing is embedded in the spatial structures.
 The time dependence of each variable
 is expressed in a given power
 of the evolution function,
 which is solved consistent to similarity.
 This evolution function $y(t)$
 agrees with the well established
 parametric solution of Mestel \citep{mestel1965}.
 The spatial structure in the Lagrangian coordinate $\eta$
 is described by an equation
 similar to that derived by Goldreich and Weber
 \citep{goldreich1980}
 with solutions covering the central core and the external envelope.
 In particular, the spatial solutions generate configurations
 that are relevant to planetary system in protostar
 and cavity in pre-supernova.
 The most important points of this Lagrangian description
 are that the evolution function
 is not limited to a linear time scaling,
 and the spatial solutions appear to be compatible to physical systems.
 These factors greatly enhance the scope of homologous treatment.
 We have the believe that the Lagrangian homologous description
 might be the mathematical tool to describe
 some self-organized astrophysical phenomena.

\acknowledgments

\appendix

\clearpage
\begin{figure}
\plotone{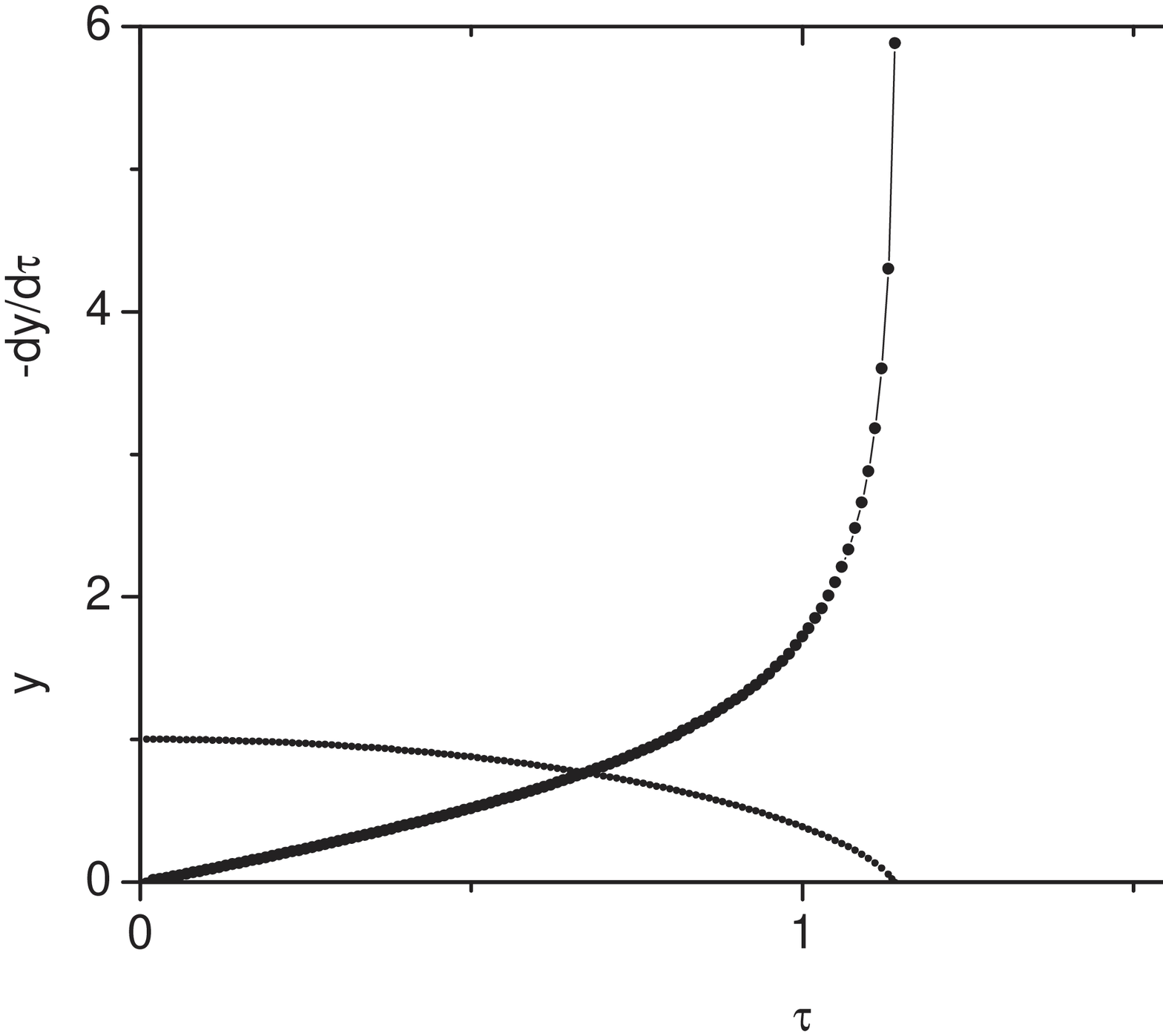}
\caption{The evolution function $y$ in thin line
 and its negative time derivative $-dy/d\tau$ in thick line
 are plotted as a function of the normalized time $\tau$.}
\label{fig.1}
\end{figure}

\clearpage
\begin{figure}
\plotone{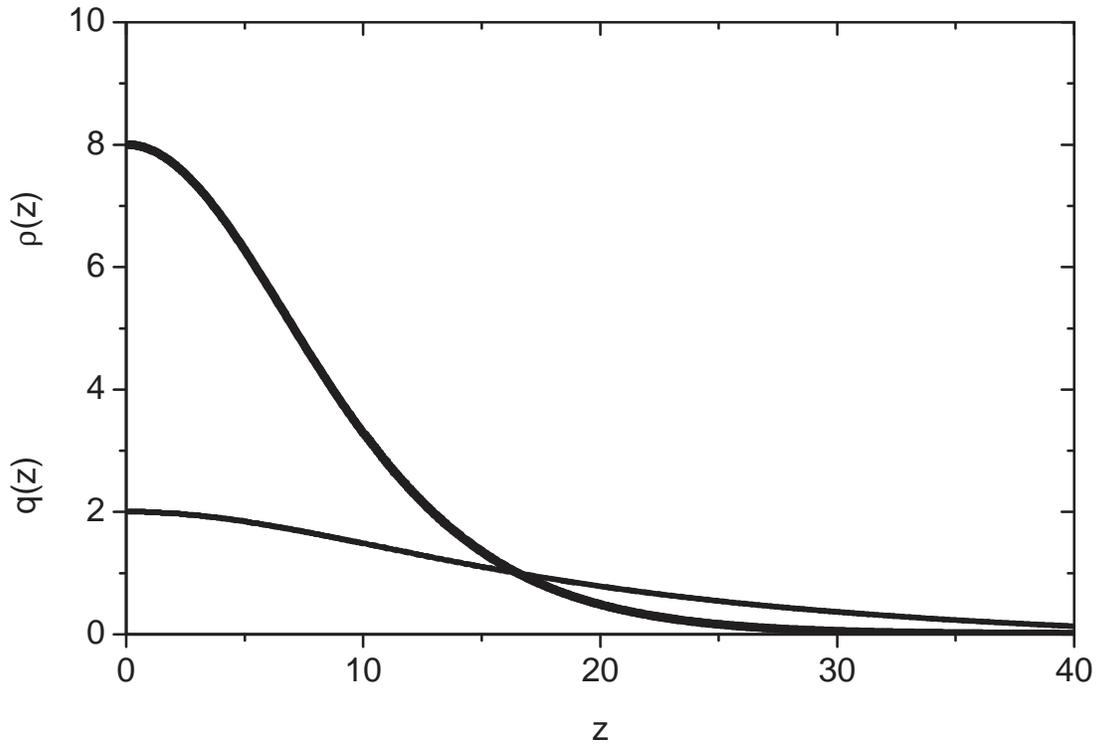}
\caption{The equilibrium profile of
 density $\bar\rho=q^{3}$ in thick line
 and $q$ in thin line
 are shown as a funtion of the normalized distance $z$
 with $q(0)=2$ and $\alpha=100$ giving $z_{max}=40$.}
\label{fig.2}
\end{figure}

\clearpage
\begin{figure}
\plotone{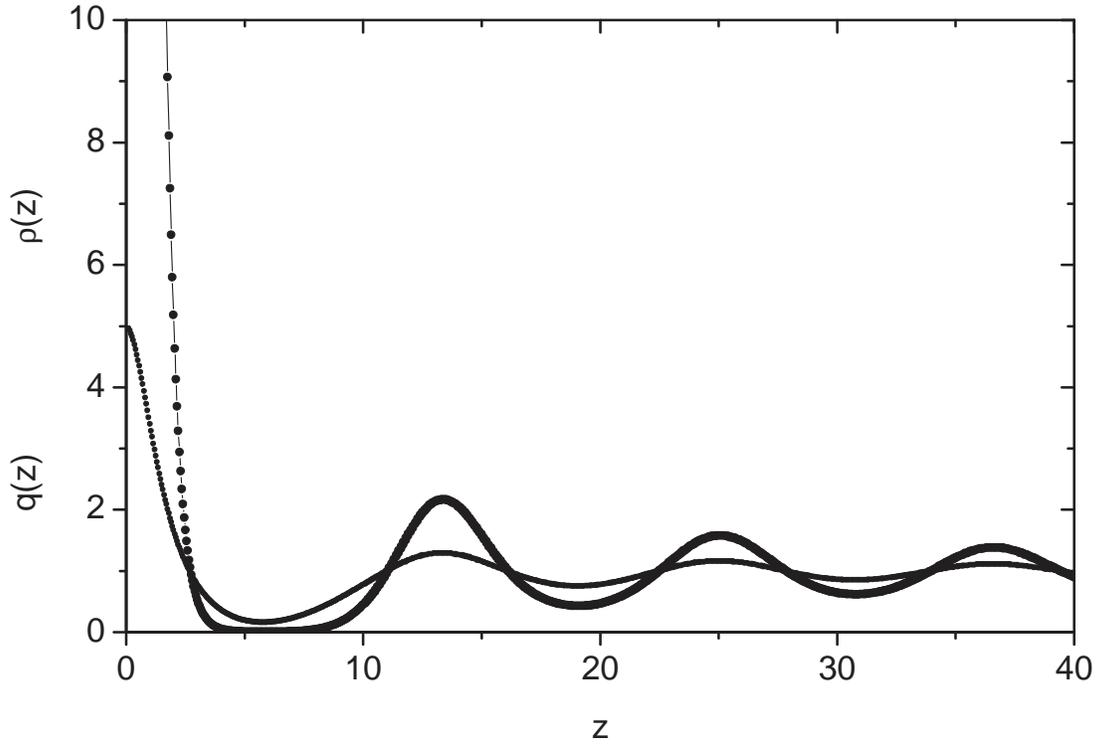}
\caption{The collapsing profile of
 density $\bar\rho=q^{3}$ in thick line
 and $q$ in thin line
 are shown as a funtion of the normalized distance $z$
 with $q(0)=5$ and $\alpha=5$.
 The peak density around $z=0$ is off scale.}
\label{fig.3}
\end{figure}

\clearpage
\begin{figure}
\plotone{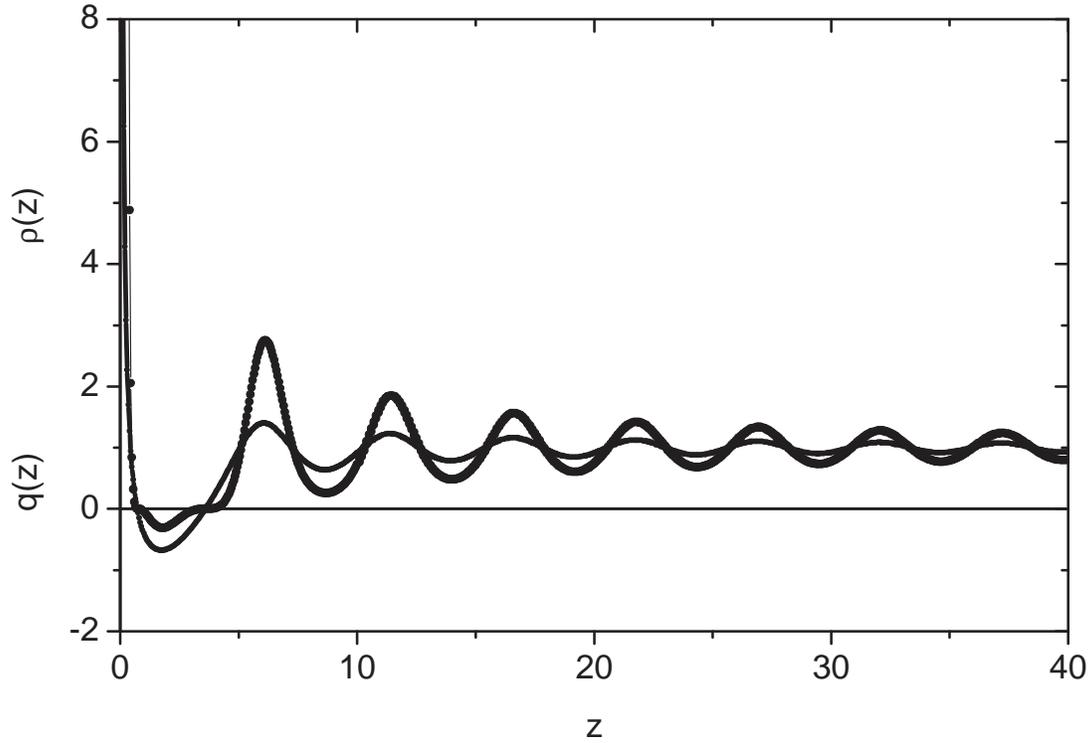}
\caption{The collapsing profile of
 density $\bar\rho=q^{3}$ in thick line
 and $q$ in thin line
 are shown as a funtion of the normalized distance $z$
 with $q(0)=20$ and $\alpha=1$ giving $z_{0}=4.5$.
 The peak $\bar\rho$ and $q$ around $z=0$ are off scale.}
\label{fig.4}
\end{figure}

\clearpage
\begin{figure}
\plotone{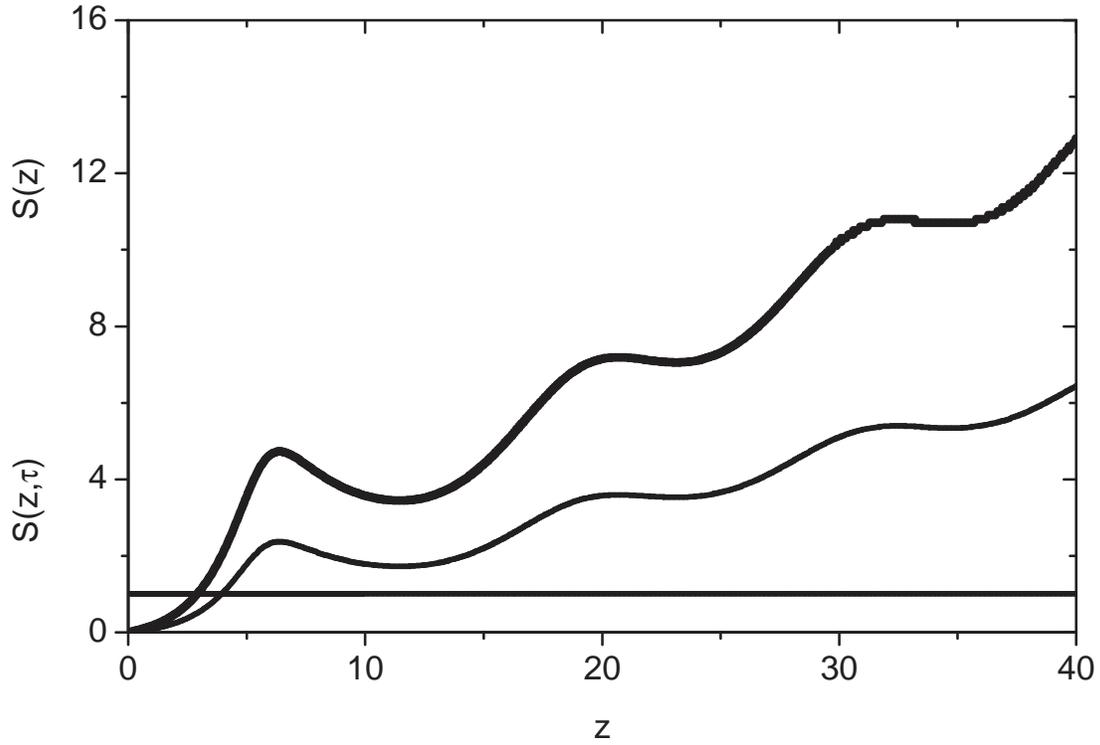}
\caption{The spatial part of the sonic function $\bar S(z)$ in thick line
 and the time dependent sonic function $S(z,\tau)$ in thin line
 at the moment of $dy/d\tau=0.5$
 are plotted as a funtion of the normalized distance $z$
 with $q(0)=5$ and $\alpha=5$.
 The sonic surface is at $z_{s}=3.95$.}
\label{fig.5}
\end{figure}

\end{document}